\documentclass[conference]{IEEEtran}
\usepackage{cite}

\ifCLASSINFOpdf
  \usepackage[pdftex]{graphicx}
\else
  \usepackage[dvips]{graphicx}
\fi
\usepackage[cmex10]{amsmath}
\usepackage[noend]{algorithmic}

\newsavebox{\ieeealgbox}
\newenvironment{boxedalgorithmic}
  {\begin{lrbox}{\ieeealgbox}
   \begin{minipage}{\dimexpr\columnwidth-2\fboxsep-2\fboxrule}
   \begin{algorithmic}[1]
   }
  {
   \end{algorithmic}
   \end{minipage}
   \end{lrbox}\noindent\fbox{\usebox{\ieeealgbox}}}

\usepackage{array}
\usepackage{url}

\begin{document}
%
\title{Triggered Clause Pushing for IC3}

\author{\IEEEauthorblockN{Martin Suda
}
\IEEEauthorblockA{
Max-Planck-Institut f\"ur Informatik, Saarbr\"ucken, Germany}
\IEEEauthorblockA{
Saarland University, Saarbr\"ucken, Germany}
\IEEEauthorblockA{
Charles University, Prague, Czech Republic}
}


%


\maketitle


\begin{abstract}
We propose an improvement of the famous IC3 algorithm for model checking safety properties of finite state systems.
We collect models computed by the SAT-solver during the clause propagation phase of the algorithm
and use them as witnesses for why the respective clauses could not be pushed forward.
It only makes sense to recheck a particular clause for pushing when its witnessing model falsifies a newly added clause.
Since this trigger test is both computationally cheap and sufficiently precise,
we can afford to keep clauses pushed as far as possible at all times.
Experiments indicate that this strategy considerably improves IC3's performance.
\end{abstract}


%
\IEEEpeerreviewmaketitle




\section{Introduction}

IC3 \cite{DBLP:conf/vmcai/Bradley11} is one of the strongest 
 bit-level safety model checking algorithms currently known.
Its highly focused reasoning guided by the property being analyzed results
in remarkable performance in proving safety
complemented by a unique ability to find deep counterexamples.
By leveraging the power of modern incremental SAT-solvers 
IC3 carefully updates clausal reachability information
while maintaining a surprisingly small memory footprint.

Since its discovery by Aaron Bradley in 2010, IC3 has drawn attention of many researchers.
It has been extended to deal with liveness properties \cite{DBLP:conf/fmcad/BradleySHZ11},
applied to incremental verification \cite{DBLP:conf/fmcad/ChocklerIMMN11},
and generalized to model checking software \cite{DBLP:conf/cav/CimattiG12}.
An inspiring paper by E{\'e}n et al.~\cite{DBLP:conf/fmcad/EenMB11}
presents a detailed account of an efficient implementation of IC3
and advocates the importance of studying the algorithm further.

In this paper we focus on the so-called propagation phase of the algorithm,
where the clauses learned so far are inspected to check whether they
could yield an inductive invariant proving the property. This is done
by attempting to ``push''  individual clauses forward in an operation
involving a specific query to a SAT-solver. Normally,
if the query is satisfiable, the clause cannot be pushed
and the derived model is thrown away. We propose to keep the model instead
and use it as a witness for why the clause cannot be pushed.
The key observation is that it only makes sense to repeat the expensive
SAT-solver call when the witness has been subsumed by another clause.

Being equipped with this cheap trigger test 
allows us to incorporate clause propagation directly into the main loop of the algorithm.
A clause is pushed forward as soon as its context becomes strong enough
to make the above query unsatisfiable.
This provides IC3 with a better guidance and enables immediate
detection of convergence to the invariant.

Our experiments show that using the witnesses pays off in practice.
Moreover, they provide opportunity for further refinements of the algorithm.
We present a new clause minimization heuristics aimed at subsuming
as many witnesses as possible.

The rest of the paper is structured as follows. 
After fixing the terminology in Section \ref{sec_prelim},
we give an overview of IC3 in Section \ref{sec_overview}.\footnote{
Due to space limitations, some aspects of the algorithm 
could not be presented in full detail. If necessary, 
we recommend the reader to consult the original work \cite{DBLP:conf/vmcai/Bradley11}
or the paper \cite{DBLP:conf/fmcad/EenMB11},
from which we adopt some of the notation.}
Triggered pushing is explained in detail in Section~\ref{sec_pushing}
and incorporated into the overall algorithm in Section~\ref{sec_subsumption}.
We also show there how exhaustive subsumption can be performed
efficiently in IC3, which may be of independent interest.
We presents our experiments in Section~\ref{sec_experiment}
and conclude in Section~\ref{sec_discuss} with final remarks.


\section{Preliminaries}

\label{sec_prelim}

We assume the system to be verified is modeled as a finite state machine (FSM).
A FMS $M = \langle X,I,P,T \rangle $ is described by a finite set of Boolean state variables $X$,
such that each assignment $s \in \{0,1\}^X$ corresponds to a \emph{state} of $M$,
 further 
by sets of \emph{initial} $I \subseteq \{0,1\}^X$ and \emph{property} $P\subseteq \{0,1\}^X$ states, 
and by a \emph{transition} relation $T \subseteq \{0,1\}^X \times \{0,1\}^X$.
States not satisfying the property $P$ are referred to as \emph{bad} states.
A path in $M$ is a sequence $s_0,\ldots,s_k$ of states, such that 
$\langle s_i,s_{i+1}\rangle \in T$ for every $0 \leq i < k$.
The model checking algorithm establishes whether there exists
a path from an initial state to a bad state.
The system is deemed \emph{safe} if no such \emph{counterexample} path exists.
Safety may be shown by providing an \emph{inductive invariant} proving $P$,
which is a formula $\varphi$, such that $I \Rightarrow \varphi$,
$\varphi \land T \Rightarrow \varphi'$, and $\varphi \Rightarrow P$.
Here we use the convention that priming a formula means interpreting
it over the next state variables.

A \emph{literal} is a state variable or its negation. 
A consistent conjunction of literals is referred to as a \emph{cube}
and a disjunction as a \emph{clause}.
A set of clauses stands for their conjunction.
States of the FSM naturally correspond to cubes mentioning every variable from $X$.
The FSM is assumed to be symbolically represented in a such way that allows us 
to employ a SAT-solver for answering queries about it. 


\renewcommand{\algorithmicrequire}{\textbf{Algorithm}} 

\begin{figure}[!t]
\algsetup{indent=1em}
\begin{boxedalgorithmic}
\REQUIRE{IC3 (FSM $M = \langle X,I,P,T \rangle $)}
\STATE $L_0 \leftarrow I$; \textbf{foreach} $j > 0: L_j \leftarrow \emptyset$
\FOR{$k=0,1,\ldots$}
	\STATE \COMMENT{Recursive path construction / blocking}
	\WHILE{$\mathit{SAT?}[L_k \land \neg P]$} \label{ic3_line_sat1}
		\STATE extract state $s$ from the model \label{ic3_line_extract_bad}
		\STATE $\mathit{Q} \leftarrow \{ \langle s,k \rangle \}$ \label{ic3_line_q_init}
		\WHILE{$Q$ \NOT empty} \label{ic3_line_blockingcycle}
		\STATE pop some $\langle s,i \rangle$ from $Q$ with minimal $i$ \label{ic3_line_pickclose}
		\IF{$i = 0$}
			\RETURN COUNTEREXAMPLE \label{ic3_line_cex}
		\ENDIF
		\IF{$\mathit{SAT?}[L_{i-1} \land T \land s']$} \label{ic3_line_sat2}
			\STATE extract a predecessor state $t$ from the model \label{ic3_line_extract_model}
			\STATE $\mathit{Q} \leftarrow Q \cup \{ \langle t,i-1\rangle, \langle s,i\rangle \}$ \label{ic3_line_storeboth}
		\ELSE
			\STATE extract the used assumptions $s_0' \subseteq s'$ \label{ic3_line_assumps}
			\STATE \textbf{foreach} $0 \leq j \leq i:$ $L_j \leftarrow L_j \cup \{\neg s_0\}$  \label{ic3_line_strengthen}
			\IF{$i < k$} \label{ic3_line_resched1}
				\STATE $\mathit{Q} \leftarrow Q \cup \{ \langle s,i+1\rangle \}$ \label{ic3_line_resched2}
			\ENDIF
		\ENDIF
		\ENDWHILE
	\ENDWHILE
		
	\STATE
	\STATE \COMMENT{Clause propagation}
	\FOR{$i=0,\ldots,k$ and \textbf{foreach} $c \in L_i \setminus L_{i+1}$} \label{ic3_line_pushstart}
		\IF{\NOT $\mathit{SAT?}[L_i \land T \land \neg c']$} \label{ic3_line_sat3}
			\STATE $L_{i+1} \leftarrow L_{i+1} \cup \{c\}$
		\ENDIF
		\IF{$L_i = L_{i+1}$} \label{ic3_line_repcheck}
		\RETURN SAFE \label{ic3_line_proven}
		\ENDIF
	\ENDFOR
\ENDFOR
\end{boxedalgorithmic}
\caption{\emph{High-level description of IC3.} Some features of the algorithm
not relevant for our presentation have been omitted. Please consult \cite{DBLP:conf/vmcai/Bradley11}
on how to strengthen the query on line \ref{ic3_line_sat2} with induction and \cite{DBLP:conf/fmcad/EenMB11} on how to generalize 
states on lines \ref{ic3_line_extract_bad} and \ref{ic3_line_extract_model} with ternary simulation.
}  
\label{ic3_pseudo}
\end{figure}



\section{Overview of IC3}

\label{sec_overview}

The IC3 algorithm can be seen as a hybrid between 
explicit and symbolic approach to model checking.
It explicitly constructs a path, starting from a
bad state and extending it backwards towards an initial state.
At the same time, it maintains symbolic stepwise approximating
reachability information, which is locally refined
whenever the current path cannot be extended further.
The reachability information guides the path construction,
and is also bound to eventually converge to a proof of safety,
if no full path exists.

Specifically, IC3 maintains a sequence of sets of clauses
$L_0, L_1, \ldots$, which we call \emph{layers}. 
Layers are updated in an iterative manner, such
that they satisfy the following properties:
1) $L_0 \equiv I$,
2) $L_i \supseteq L_{i+1}$ and thus $L_i \Rightarrow L_{i+1}$ for every $i$,
3) 
 $L_{i+1}$ is an overapproximation of the image of $L_i$ for every $i$,
4) at the end of iteration $k$ of the algorithm
there is no bad state satisfying $L_k$.
It follows that on successful termination of iteration $k$,
IC3 will have established that
there is no counterexample path of length $k$ or less.

Let us now have a look at the pseudocode of IC3 in \figurename~\ref{ic3_pseudo}.
We see that initially $L_0$ is identified with $I$\footnote{
We assume here that $I$ has a feasible description as a set of clauses over $X$.
Indeed, it is typically translated into a set of unit clauses.
Minor changes are needed (see \cite{DBLP:conf/fmcad/EenMB11}) to accommodate to the general case.
} 
and all the other layers are empty. 
Each iteration then comprises two phases:
a blocking phase and a propagation phase.
The \emph{blocking} phase maintains a set $Q$,
working as a priority queue, of so-called \emph{proof obligations},
pairs of the form $\langle s,i \rangle$, where $s$ is a state 
that can reach a bad state and $i$ is an index.
Successfully blocking a proof obligation $\langle s,i \rangle$
amounts to showing that $s$ cannot reach an initial state in
at most $i$ steps. Such information is recorded as a new clause
strengthening the layer $L_i$. 
Deriving this clause may require first recursively blocking
other obligations, corresponding to predecessor states of $s$,
and strengthening the previous layers.

The blocking phase of iteration $k$ starts by using a SAT-solver to pick
a bad state $s$ satisfying $L_k$ (lines~\ref{ic3_line_sat1} and \ref{ic3_line_extract_bad}).
Then the set $Q$ is initialized for blocking the obligation $\langle s,k \rangle$ (line~\ref{ic3_line_q_init}).
The inner loop (starting at line~\ref{ic3_line_blockingcycle}) processes individual obligations
picking first those that are estimated to be closer to an initial state (line~\ref{ic3_line_pickclose}).
An obligation with $i=0$ means a full counterexample path has been constructed 
and the algorithm terminates (line~\ref{ic3_line_cex}).
If the SAT-solver query on line~\ref{ic3_line_sat2} returns SAT,
we extract a predecessor state $t$ known to satisfy $L_{i-1}$.
This signifies progress in extending the current path from $s$ to $t$,
or, equivalently, a current failure to block the obligation $\langle s,i \rangle$.
Both the new obligation $\langle t,i-1 \rangle$ to be worked on next
and the current are stored in $Q$ (line~\ref{ic3_line_storeboth}).
If, on the other hand, the above call returns UNSAT, 
we assume the solver provides us with a subset $s_0$ of the state assumptions $s$
that were needed in the proof (line~\ref{ic3_line_assumps}). 
This corresponds to generalizing the reason for why the obligation
was blocked. The obtained subset $s_0$ understood as a cube, 
becomes a clause 
when negated by which the algorithm strengthens the layers $L_0,\ldots,L_i$ (line~\ref{ic3_line_strengthen}).\footnote{
For efficiency, $s_0$ should be as small as possible to provide for a good generalization.
For correctness, $s_0$ must not intersect $I$. This can always be achieved,
since at this point the state $s$ is never an initial state.
} Finally, the blocked obligation $\langle s,i \rangle$ may be \emph{rescheduled}
by one step (lines \ref{ic3_line_resched1}, \ref{ic3_line_resched2}).\footnote{
These two lines are not needed for correctness, but they substantially improve IC3's performance.
When left out, the set $Q$ operates as a stack and
forces IC3 to find counterexamples of minimal length.}

Let us now turn to the \emph{propagation} phase, which follows next (starting on line \ref{ic3_line_pushstart}).
It scans the layer clauses one by one and checks with the help of a SAT-solver call (line \ref{ic3_line_sat3}) 
for each $c \in L_i \setminus L_{i+1}$ whether it can be ``pushed'' to strengthen a layer with a higher index.
The clause is successfully pushed forward when the solver returns UNSAT, having proved that $L_i \land T \Rightarrow c'$.
If it is detected during propagation that two neighboring layers have been made identical (line \ref{ic3_line_repcheck}),
the algorithm terminates reporting that no counterexample is possible (line \ref{ic3_line_proven}).
The justification for this conclusion follows from the four properties of
layers mentioned earlier. The repeating layer $L_i$, in fact,
forms an inductive invariant which proves the system to be safe.



\section{Triggered Clause Pushing}

\label{sec_pushing}


There are several reasons for why the clause propagation phase 
is an important part of IC3. First, it is an opportunity
to insert clauses into the till now empty layer $L_{k+1}$ before the start of iteration $k+1$.
Sometimes, thanks to pushed clauses, iterations pass off without actually entering the blocking loop.
Second, it generally strengthens the layers
which then provide better guidance for path construction or,
equivalently, a stronger context for obligation blocking.
Finally, and most importantly, clause propagation is the place where 
the algorithm's convergence to an inductive invariant is detected.

For these reasons it could be advantageous to perform
clause propagation more often than just once per iteration.
There is, however, a non-trivial computational cost connected with propagation,
and so it can only pay off to run it again when the layers have changed sufficiently since it was last performed.
Here we show how to detect on a per clause basis that a previously failed 
pushing attempt should be reconsidered. This will allow us 
to come up with a version of IC3, where all the clauses are pushed
as far as possible at all times.

Consider a clause $c \in L_i \setminus L_{i+1}$ that could not be pushed forward.
This means the query on line \ref{ic3_line_sat3} 
of the pseudocode in \figurename~\ref{ic3_pseudo} returned SAT.
We may now inspect the model computed by the SAT-solver
and extract a state $w_c$ which satisfies $L_i$
and from which there is a transition to a state satisfying $\neg c$.
Notice that as long as $w_c$ remains to satisfy $L_i$
during the potential strengthenings of the layer, the query in question cannot become UNSAT.
The state $w_c$, therefore, represents a \emph{witness} for why $c$
cannot be pushed forward from $L_i$ to $L_{i+1}$.	


But how do we efficiently recognize whether $w_c$ still satisfies $L_i$
after a new clause $d$ has been added to $L_i$? The answer is: via subsumption!
It is only when $d \subseteq \neg w_c$ (here we again use that negation of a cube is a clause)
that $w_c$ ceases to be a witness, because it does not satisfy the strengthened $L_i$.
Now we may directly retry the pushing query of line \ref{ic3_line_sat3}
and either discover a new witness or finally push the clause $c$ to $L_{i+1}$.

It may seem expensive to perform the subsumption test against every witness
whenever a new clause is derived. Note, however, that efficient implementations 
of IC3 already use subsumption routinely to test each new clause against all other clauses
(subsumed clauses can be removed which helps to keep the layers small)
and that by also considering the witnesses, one per each clause, the overhead
is at most doubled. In the next section we explain how to exploit the semantic relation
between the individual layers to potentially reduce this cost.

\section{IC3 with subsumption and triggered pushing}

\label{sec_subsumption}

It has been observed that IC3 often derives a clause $c$ to be inserted into layer $L_i$
while $L_i$ already contains a weaker clause $d \supseteq c$.
Bradley \cite{DBLP:conf/vmcai/Bradley11} proposes to remove such clauses during propagation;
the implementation described in \cite{DBLP:conf/fmcad/EenMB11} is more eager
and clears layers via subsumption each time a new clause is derived.
Removing subsumed clauses pays off, because they do not bring any additional information
only make the layers unnecessarily large. 


\begin{figure}[!t]
\centering
\includegraphics[trim=24 0 0 0, clip]{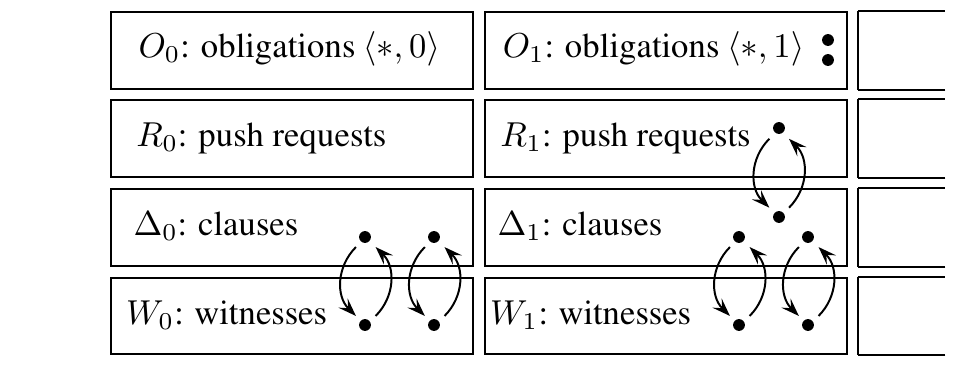}
\vspace{-15pt} 
\caption{\emph{Organizing the data structures of IC3 with triggered pushing.} 
A bi-directional link is maintained between a clause and its witness / push request.}
\label{fig_organize}
\end{figure}

Once subsumption is implemented for reducing layers it can also be used
for pruning proof obligations. Indeed, by construction, the clause $c$
learned while blocking an obligation $\langle s,i \rangle$ satisfies
$c \subseteq \neg s$, but may also subsume other obligations
$\langle t,i \rangle$ currently on $Q$. These can be directly rescheduled
to index $i+1$, each saving us one SAT-solver call.

Now we describe how to organize the data structures of IC3 
such that 1) subsumption by newly derived clauses
can be used to prune layers and obligations, 2)
clause pushing triggered by subsuming a witness
is integrated into the blocking phase to keep 
clauses pushed as far as possible at all times.

To avoid duplicating clauses we use the delta encoding of layers
proposed in \cite{DBLP:conf/fmcad/EenMB11}. A delta layer $\Delta_i$
consist of clauses appearing last in $L_i$. 
Thus $\Delta_i = L_i \setminus L_{i+1}$ and $L_i = \bigcup_{j \geq i} \Delta_j$.
Each layer clause $c$ is either associated with its witness $w_c$
or a \emph{push request} is stored for it, which means 
it will need to be considered for pushing.
Finally, instead of using a priority queue, we
explicitly separate proof obligations into sets $O_i$ based on their index.
The whole situation is depicted in \figurename~\ref{fig_organize}.

The algorithm now works as follows. It picks the smallest index $i$
such that there is either an obligation in $O_i$ or a request in $R_i$.
If both sets are non empty, obligations are picked first.\footnote{
First, by blocking obligations from $O_i$ we strenghten $L_i$.
Then we consider the requests from $R_i$.
If a clause is successfully pushed to $L_{i+1}$
it may subsume obligations waiting in $O_{i+1}$.}
Handling a proof obligation corresponds to 
asking the query from line \ref{ic3_line_sat2} in \figurename~\ref{ic3_pseudo}
and either creates a new obligation or derives a new clause to be added to $\Delta_i$.
Similarly, handling a push request corresponds to the query of line \ref{ic3_line_sat3}
and either generates a new witness, which is stored to $W_i$, 
or pushes the clause from $\Delta_i$ to $\Delta_{i+1}$.
In both cases a new clause may be added to a layer,
which is where subsumption comes into play.

When a clause $c$ is added into $\Delta_i$
we put a push request for it into $R_i$ and then
do the following: 1) we remove all the clauses from $\Delta_i$
subsumed by $c$ (along with their witnesses or associated push requests),
2) we remove the subsumed witnesses from $W_i$ and 
insert push requests for the respective clauses into $R_i$,
3) we reschedule the subsumed proof obligations from $O_i$ to $O_{i+1}$.
If the clause $c$ was pushed to $\Delta_i$ from $\Delta_{i-1}$,
we are done. If, on the other hand, $c$ was derived
during blocking, it formally strengthened all $L_0,\ldots,L_i$.
We, therefore, continue towards lower indices performing
1) and 2)\footnote{The sets $O_j$ of proof obligations
are empty for $j < i$ at this point. 
} for $j=i-1,i-2,\ldots$ A key observation is
that the iteration can be stopped as soon as the clause $c$ 
is itself subsumed by some clause $d$ from $\Delta_j$. 
Since layers of low index are stronger than those further on,
the iteration typically terminates way before reaching $j=0$.
This way a lot of time spent on futile subsumption tests can be saved.


\section{Experiment}

\label{sec_experiment}

To experimentally evaluate the benefit of the presented technique
we implemented both the standard IC3 algorithm and its variation
extended with triggered pushing and compared them on the benchmarks
from the Hardware Model Checking Competition of 2012.\footnote{See \url{http://fmv.jku.at/hwmcc12/}.}
Since most of the code is shared by the two implementations
the results should directly reflect the relative improvement caused by triggered pushing
which is expected to carry over to other implementations.


\begin{figure}[!t]
\centering
\includegraphics[scale=1.1]{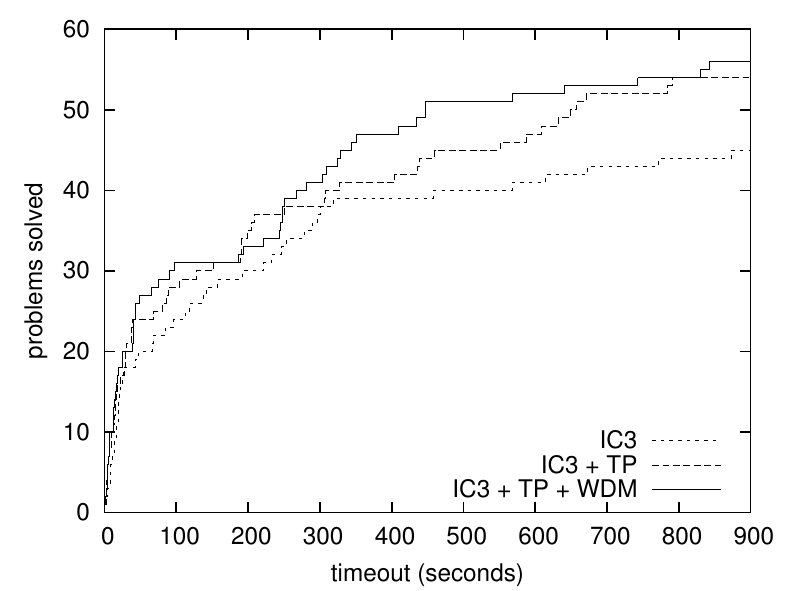}
\vspace{-20pt} 
\caption{Comparing original IC3 to a version with triggered pushing (TP) and to one further enhanced by witness directed minimization (WDM).}
\label{fig_compare_orig}
\end{figure}

Our code\footnote{Available at \url{http://www.mpi-inf.mpg.de/~suda/triggered.html}.} is built on top of the SAT-solver Minisat \cite{DBLP:conf/sat/EenS03} version 2.2.
We transform the circuit to CNF using the Plaisted-Greenbaum encoding \cite{DBLP:journals/jsc/PlaistedG86}
which is then simplified by variable elimination \cite{DBLP:conf/sat/EenB05}.
The obligation queue and layers are organized as described in Section~\ref{sec_subsumption}.
We found it advantageous in this setup to allocate a new solver instance for
every time index. That way, the solvers corresponding to strong layers of low indices
are not polluted by the weaker clauses derived further on. Clauses, as well as
states of proof obligations and witnesses, are stored sorted which enables
a linear pass subsumption test. The test is, however, only started if 
the inputs pass a pre-filter based on precomputed signatures \cite{DBLP:conf/sat/EenB05}.

Before we present our experimental results let us explain one further enhancement of IC3
which is readily available once the witnesses for pushing are maintained. 
Recall that when a proof obligation is succcessfully blocked 
a set of used assumptions is extracted from the SAT-solver.
It is important for efficiency that this set be as small as possible.
That is why this set is usually explicitly minimized 
by removing individual literals and checking whether the respective query remains UNSAT.
It has been observed \cite{DBLP:conf/vmcai/Bradley11} that the order in which literals are tried for removal
affects the quality of the final result. Here we propose 
a heuristical order aimed at subsuming witnesses and thus evoking pushing:
a particular literal is preferred for removal when
there is a high number of witnesses of the respective layer 
that would not be subsumed if the literal remained in the learned clause.
The idea is that early removals are easier then later ones and
so with this order we try to keep the chance of subsuming a witness
by the learned clause high. We call the technique witness directed minimization (WDM).

Let us finally have a look at \figurename~\ref{fig_compare_orig}, which compares
the performance of the original IC3, a version with triggered pushing,
and a version further extended with WDM.
(The first two versions use a random literal order for clause minimization.)
We ran the versions separately on our servers with 3.16 GHz Xeon CPU, 16 GB RAM, and Debian 6.0.
The timeout was set to 900 seconds per problem.
In the end original IC3 solved 45 problems (19 SAT and 26 UNSAT),
a version with triggered pushing 54 (18 SAT and 36 UNSAT),
and the one with WDM 56 problems (18 SAT and 38 UNSAT). This clearly demonstrates
that triggered pushing considerably improves the performance of IC3.





\section{Discussion}

\label{sec_discuss}

Out of curiosity, we performed the above experiment with a version
of IC3 completely without clause propagation where convergence is detected when 
a delta layer becomes empty due to subsumption. This version solved 19 problems (15 SAT and 4 UNSAT).
This experimentally confirms that clause propagation is an important phase
during which IC3 strengthens its layers to provide better guidance for subsequent iterations
and, more importantly, establishes whether convergence to an inductive invariant has occurred.

It this paper we have shown how the power of clause propagation
can be directly incorporated into the main loop of IC3 rendering its benefits continuous.
This is done by leveraging witnesses, states extracted from failed clause pushing
attempts, which would normally be thrown away. Maintaining the witnesses
provides new opportunities for directing IC3 towards the invariant,
as exemplified by the witness directed minimization technique we proposed.
We believe the witnesses could also be used as a theoretical tool 
for a deeper understanding of the remarkable performance of IC3 in general.

\bibliographystyle{IEEEtran}
\bibliography{IEEEabrv,triggeredPushing}

\end{document}